\documentclass[a4paper,conference,10pt]{IEEEtran}

\usepackage{amssymb}
\usepackage{amsmath}
\usepackage{amsthm}
\usepackage{amsfonts}
\usepackage{accents}
\usepackage{multirow}
\usepackage{array}
\usepackage{multirow}
\usepackage{graphicx}
\usepackage{epstopdf}
\usepackage{float}
\usepackage{pgfplots}
\pgfplotsset{compat=1.8}
\usepackage{epstopdf,epsfig}

\usepackage[switch]{lineno}
\usepackage[named]{algo}
\usepackage[noend]{algpseudocode}
\usepackage{algorithm}

\ifCLASSOPTIONcompsoc
  \usepackage[caption=false,font=normalsize,labelfont=sf,textfont=sf]{subfig}
\else
  \usepackage[caption=false,font=footnotesize]{subfig}
\fi

\usepackage{balance}

\usepackage{booktabs}

\definecolor{mycolor1}{rgb}{0.00000,0.20000,0.60000}%
\definecolor{mycolor2}{rgb}{0.00000,0.40000,0.80000}%
\definecolor{mycolor3}{rgb}{0.40000,0.60000,1.00000}%
\definecolor{mycolor4}{rgb}{0.00000,0.60000,1.00000}%
\definecolor{mycolor5}{rgb}{1.00000,0.40000,0.60000}%
\definecolor{mycolor6}{rgb}{1.00000,0.20000,0.40000}%
\definecolor{mycolor7}{rgb}{1.00000,0.00000,0.20000}%
\definecolor{mycolor8}{rgb}{0.60000,0.20000,0.00000}%

\DeclareMathOperator{\T}{\text{T}}
\DeclareMathOperator{\E}{\mathbb{E}}

\newtheorem{theorem}{Theorem}

\newtheorem{proposition}[theorem]{Proposition}

\newtheoremstyle{colon}%
{}
{}
{\rm}
{}
{\itshape}
{:}
{ }
{\thmname{#1}\thmnumber{ \itshape#2}\thmnote{ (#3)}}

\theoremstyle{colon}
\begin{document}
\title{Study of Clustered Robust Linear Precoding for Cell-Free MU-MIMO Networks}

\author{André R. Flores$^1$ and Rodrigo C. de Lamare$^{1,2}$ \\
$^1$Centre for Telecommunications Studies, Pontifical Catholic
University of Rio de Janeiro, Brazil \\
$^2$Department of Electronic Engineering, University of York,
United Kingdom \\
Emails: \{andre\_flores, delamare\}@puc-rio.br
}

\maketitle

\begin{abstract}
Precoding techniques are key to dealing with multiuser interference in the downlink of cell-free (CF) multiple-input multiple-output systems. However, these techniques rely on accurate estimates of the channel state information at the transmitter (CSIT), which is not possible to obtain in practical systems. As a result, precoders cannot handle interference as expected and the residual interference substantially degrades the performance of the system. To address this problem, CF systems require precoders that are robust to CSIT imperfections. In this paper, we propose novel robust precoding techniques to mitigate the effects of residual multiuser interference. To this end, we include a loading term that minimizes the effects of the imperfect CSIT in the optimization objective. We further derive robust precoders that employ clusters of users and access points to reduce the computational cost and the signaling load. Numerical experiments show that the proposed robust minimum mean-square error (MMSE) precoding techniques outperform the conventional MMSE precoder for various accuracy levels of CSIT estimates.
\end{abstract}


%

\section{Introduction}
Cellular networks today rely on coordinated base stations (BSs) deployed to cover an extensive area and provide a variety of services to the users. 
Future applications of wireless communications require cellular networks to offer higher throughput \cite{Tataria2021,Giordani2020,mmimo,wence}. Since further densification of BSs is impractical, cell-free (CF) multiple-input multiple-output (MIMO) systems have emerged as an attractive alternative to improve the overall performance and support future applications \cite{Ammar2022, Elhoushy2021}. Unlike BS-based networks, CF-MIMO  employs multiple access points (APs) distributed over the area of interest. The APs are coordinated by a central processing unit (CPU) located at the cloud server. This distributed deployment has benefits such as increased throughput per user \cite{Yang2018,Elhoushy2021b,Ammar2022,mashdour2022enhanced,cl&sched} as well as better energy efficiency \cite{Ngo2018,Zhang2019,Jin2021} when compared with the conventional BS-based networks. 

To fully harness the benefits of CF-MIMO systems in the downlink and handle multiuser interference (MUI), precoding techniques \cite{joham,siprec,gbd,wlbd,cqabd,rsbd,mbthp,rmbthp,rsthp,1bitprec,rmmse,tds,itapsprec&pa,rmmsecf,rapa,cfrs,rprec&sr,lrcc,zcprec} are employed. Linear precoding such as matched filter (MF), zero-forcing (ZF) \cite{Nayebi2017}, and minimum mean-square error (MMSE) \cite{Bjoernson2020} precoders are widely used in CF MU-MIMO. Initially, these techniques were employed as network-wide (NW) precoders \cite{Ngo2017,Nayebi2017}. But
NW precoders are not practical because they involve extremely high signaling loads and computational cost. 

To address the problems with traditional NW precoders, 
several studies recommend to employ a reduced set of APs and users \cite{Bjoernson2020a,Buzzi2020,Flores2023}. Following this technique, NW precoders have evolved to precoders that employ APs and user clustering \cite{Palhares2020,Bjoernson2020a,Lozano2021} to address the high computational complexity of NW precoders. In \cite{Palhares2020}, the number of APs is reduced to make better use of the resources and diminish the signaling load. Scalable MMSE  
combiners are derived along with precoders by exploiting the uplink-downlink duality in \cite{Bjoernson2020a}. A regularized ZF precoder is proposed in \cite{Lozano2021}, where several subsets are formed to reduce the number of APs serving each user. In \cite{Flores2022}, clusters of users and APs are formed to reduce the overall computational complexity. 

Another major problem with precoding techniques is that they assume perfect knowledge of the channel state information at the transmitter (CSIT). In practice, such assumption is hardly verified and residual multiuser interference degrades the performance of precoding techniques \cite{Vu2007}. As a result, there is a demand for robust precoding techniques that are tilerant of imperfect CSIT. Robust precoders were initially developed in sensor array signal processing, wherein some well-established methods included diagonal loading \cite{Elnashar2006}, generalized loading \cite{Besson2005}, and worst case optimization \cite{Vorobyov2003}. Moreover, a non-conventional transmit scheme known as rate-splitting has been developed to enhance the robustness of the system,\cite{Joudeh2016,Flores2022a}. 
 A low complexity robust precoder with per antenna power constraint was proposed in \cite{Medra2018}. Furthermore, a robust MMSE precoding scheme with power allocation was developed in \cite{rmmse}. However, the computational cost of the technique in \cite{rmmse} is high because it employs NW precoders and requires searching the precoder parameter. 

In this paper, contrary to previous studies, we propose a robust MMSE precoders to mitigate the negative effects of MUI through  alternating optimization of the precoder and the regularization parameter. 
We also introduce sparse and cluster-based robust precoders to enhance the overall performance while keeping the computational cost low. In both cases, we perform AP selection to reduce the amount of signaling required. The cluster-based robust precoders employ clusters of users to reduce the computational load. Numerical experiments show that the proposed robust techniques outperforms existing precoders.  

The rest of this paper is organized as follows. In the next section, we present the system model of the downlink of a CF MIMO. In Section III, we derive the proposed robust MMSE-based precoding algorithms. We introduce the sum-rate as a perofrmance metric in Section IV and validate our methods through numerical experiments in Section V. We conclude in Section VI.  

Throughout this paper, we reserve the bold lowercase, bold uppercase, and calligraphic letters for vectors, matrices, and sets, respectively; the notations $(\cdot)^{\text{T}}$, $(\cdot)^H$, $(\cdot)^{*}$, $\lVert\cdot\rVert$ and $|\cdot|$ stand for the transpose, Hermitian, complex conjugate, Euclidean norm, and magnitude, respectively; $\textrm{Tr}(\cdot)$ and $\mathbb{E}\left[\cdot\right]$ represent trace and statistical expectation operators, respectively. The operator $\Re\left\lbrace \cdot \right\rbrace$ retains the real part of a complex argument. An $N \times K$ matrix with column vectors $\mathbf{a}_1$, $\cdots$, $\mathbf{a}_K$, each of length $N$, is $\mathbf{A} = \left[\mathbf{a}_1,\cdots,\mathbf{a}_K\right]$. 

\section{System Model}
Consider the downlink of a CF MIMO system operating with $N$ APs distributed over the area of interest. The system provides service to a total of $K$ users, each one equipped with a single omnidirecitonal antenna. A central processing unit (CPU) located at the cloud server is connected to the APs. The data are transmitted over a flat-fading channel $\mathbf{G}^{\text{T}}\in\mathbb{C}^{K\times N}$. The $(n,k)$-th element of matrix $\mathbf{G}$ is the channel coefficient between the $n$-th AP and $k$-th user, i.e.,  $g_{n,k}=\sqrt{\zeta_{n,k}}h_{n,k}$, where $\zeta_{n,k}$ is the large-scale fading coefficient that models the path loss and shadowing effects, and $h_{n,k}$ represents the small-scale fading coefficient. The coefficients $h_{n,k}$ are modeled as independently and identically distributed (i.i.d.) random variables with complex Gaussian distribution $\mathcal{CN}\left(0,1\right)$.

Denote the transmit signal by $\mathbf{x}\in \mathbb{C}^{N}$, which obeys the transmit power constraint  $\mathbb{E}\left[\lVert\mathbf{x}\rVert^2\right]\leq P_t$, where $\mathbb{E}[\cdot]$ denotes the statistical expectation. Then, the $K \times 1$ received signal vector is
\begin{equation}
\mathbf{y}=\mathbf{G}^{\text{T}}\mathbf{x}+\mathbf{n},
\end{equation}
where $\mathbf{n}\in \mathbb{C}^{K}$ is the additive white Gaussian noise (AWGN) that follows the distribution $\mathcal{CN}\left(\mathbf{0},\sigma_n^2\mathbf{I}\right)$. The transmitted signal is 
\begin{equation}
    \mathbf{x}=\mathbf{P}\mathbf{s}=\sum_{k=1}^{K}s_k\mathbf{p}_k,
\end{equation}
where the vector of symbols $\mathbf{s}=\left[s_1,s_2,\cdots,s_K\right]^{\text{T}}\in\mathbb{C}^{K}$ contains the information intended to the users and the precoding matrix $\mathbf{P}=\left[\mathbf{p}_1,\mathbf{p}_2,\cdots,\mathbf{p}_K\right]\in\mathbb{C}^{N\times K}$, where $\mathbf{p}_k$ is its $k$-th column, maps the symbols to the transmit antennas.

The system employs the time division duplexing (TDD) protocol and, therefore, the channels are estimated by leveraging upon the channel reciprocity property and pilot training \cite{Vu2007}. After receiving the pilots, the CPU computes the channel estimate $\mathbf{\hat{G}}^{\text{T}}=\left[\mathbf{\hat{g}}_1,\mathbf{\hat{g}}_2,\cdots,\mathbf{\hat{g}}_k\right]^{\text{T}}\in\mathbb{C}^{K\times N}$. The $(n,k)$-th element of matrix $\hat{\mathbf{G}}$ is
\begin{equation}
    \hat{g}_{n,k}=\sqrt{\zeta_{n,k}}\left(\sqrt{1+\sigma_e^2}h_{n,k}-\sigma_e\tilde{h}_{n,k}\right),
\end{equation}
where the variable $\hat{g}_{n,k}$ denotes the channel estimate between the $n$-th AP and the $k$-th user, $\tilde{h}_{n,k}$ is i.i.d $\mathcal{CN}\left(0,1\right)$ (independent from $h_{n,k}$) and models the error in the channel estimate, and $\sigma_e$ represents the quality of the channel state information (CSI). The error affecting the channel estimate $\hat{g}_{n,k}$ is denoted by $\tilde{g}_{n,k}=\sigma_e\sqrt{\zeta_{n,k}}\tilde{h}_{n,k}$, which is the $(n,k)$-th coefficient of the error matrix $\tilde{\mathbf{G}}$.

\subsection{Robust MMSE}
We consider an extended objective function which mitigates the effects of the imperfect CSIT. To this end, define the channel as $\mathbf{G}^{\text{T}}=\frac{1}{\tau}\left(\hat{\mathbf{G}}^{\text{T}}+\tilde{\mathbf{G}}^{\text{T}}\right)$, 
where $\tau=\sqrt{1+\sigma_e^2}$. Then, the received signal is 
\begin{equation}
    \mathbf{y}=\frac{1}{\tau}\hat{\mathbf{G}}^{\text{T}}\mathbf{P}\mathbf{s}+\underbrace{\frac{1}{\tau}\tilde{\mathbf{G}}^{\text{T}}\mathbf{P}\mathbf{s}}_{\boldsymbol{\Delta}}+\mathbf{n}.
\end{equation}

The optimal precoder would minimize the effect of the imperfect CSIT (e.g., by letting $\mathbb{E}\left[\lVert\boldsymbol{\Delta}\rVert^2\right]\to 0$) and perform as close as possible to the perfect CSIT case. To find such a precoder, we consider solving the following optimization problem 
\begin{align}
    \left\{\mathbf{P},f\right\}=& \text{argmin}~~\underbrace{\mathbb{E}\left[\lVert\mathbf{s}-f^{-1}\mathbf{y}\rVert^2\right]}_{T_1}+\mathbb{E}\left[\lVert\boldsymbol{\Delta}\rVert^2\right]\nonumber\\    
    &\text{subject to}~~\mathbb{E}\left[\lVert\mathbf{x}\rVert^2\right]=\text{tr}\left(\mathbf{P}\mathbf{P}^H\right)=P_t,\label{minimization problem MMSE-RB}
\end{align}
where $\text{tr}(\cdot)$ is the trace of its matrix argument and $f$ is a normalization factor that can be interpreted as an automatic gain control at the receivers \cite{Joham2005,rmmse}. The following Proposition~\ref{prop:robust1} provides the solution to this optimization problem
\begin{proposition}\label{prop:robust1}
    The robust precoder $\mathbf{P}^{\left(\text{r}\right)}$ that solves \eqref{minimization problem MMSE-RB} is 
    \begin{equation}    \mathbf{P}^{\left(\text{r}\right)}=f_{\tau}{\bar{\mathbf{P}}},
\end{equation}
where
\begin{equation}
{\bar{\mathbf{P}}}=(\underbrace{\hat{\mathbf{G}}^{*}\hat{\mathbf{G}}^{\text{T}}+\mathbb{E}\left[\tilde{\mathbf{G}}^{*}\tilde{\mathbf{G}}^{\text{T}}\right]+f^2\mathbb{E}\left[\tilde{\mathbf{G}}^{*}\tilde{\mathbf{G}}^{\text{T}}\right]+\lambda f^2_{\tau}\mathbf{I}}_{\mathbf{P}^{\left(i\right)}})^{-1}\hat{\mathbf{G}}^{*},\label{p_bar robust mmse}
\end{equation}
assuming that the inverse of $\mathbf{P}^{\left(i\right)}$ exists.

\end{proposition}

\begin{IEEEproof}
We have the mean squared error 
\begin{align}
    \mathbb{E}\left[\lVert\mathbf{s}-f^{-1}\mathbf{y}\rVert^2\right]=&\underbrace{\mathbb{E}\left[\mathbf{s}^H\mathbf{s}\right]}_{T_{11}}-f^{-1}\underbrace{\mathbb{E}\left[\mathbf{s}^{H}\mathbf{y}\right]}_{T_{12}}\nonumber\\
    &-f^{-1}\underbrace{\mathbb{E}\left[\mathbf{y}^{H}\mathbf{s}\right]}_{T_{13}}+f^{-2}\underbrace{\mathbb{E}\left[\mathbf{y}^{H}\mathbf{y}\right]}_{T_{14}}.\label{mse obejctive function}
\end{align}
Expanding the second term of \eqref{mse obejctive function} produces
\begin{align}  
T_{12}&=\frac{1}{\tau}\left[\text{tr}\left(\hat{\mathbf{G}}^{\T}\mathbf{P}\right)+\text{tr}\left(\E\left[\tilde{\mathbf{G}}^{\text{T}}\right]\mathbf{P}\right)\right].
\end{align}
Similarly, the third term of \eqref{mse obejctive function} is 
\begin{align}
    T_{13}&=\frac{1}{\tau}\left[\text{tr}\left(\mathbf{P}^H\hat{\mathbf{G}}^{*}\right)+\text{tr}\left(\mathbf{P}^H\E\left[\tilde{\mathbf{G}}^{*}\right]\right)\right].
\end{align}
Evaluating $T_{14}$ yields
\begin{align}
T_4&=\frac{1}{\tau^2}\left[\text{tr}\left(\mathbf{P}\mathbf{P}^{H}\tilde{\mathbf{G}}^{*}\hat{\mathbf{G}}^{\text{T}}\right)+\text{tr}\left(\mathbf{P}\mathbf{P}^{H}\mathbb{E}\left[\tilde{\mathbf{G}}^{*}\tilde{\mathbf{G}}^{\text{T}}\right]\right)\right]+\mathbf{R}_\mathbf{n},
\end{align}
where $\mathbf{R}_\mathbf{n}$ is the covariance matrix of the noise.
Assuming that the error in the channel estimate has zero mean, we have
\begin{align}
     T_1=&K-f_{\tau}^{-1}\left(\text{tr}\left(\hat{\mathbf{G}}^{\T}\mathbf{P}\right)+\text{tr}\left(\mathbf{P}^{H}\hat{\mathbf{G}}^{*}\right)\right)\nonumber\\
     &+f_{\tau}^{-2}\left(\text{tr}\left(\mathbf{P}^{H}\hat{\mathbf{G}}^{*}\hat{\mathbf{G}}^{\text{T}}\mathbf{P}\right)+\text{tr}\left(\mathbf{P}^{H}\E\left[\tilde{\mathbf{G}}^{*}\tilde{\mathbf{G}}^{\T}\right]\mathbf{P}\right)\right)\nonumber\\
     &+f^{-2}\text{tr}\left(\mathbf{R}_\mathbf{n}\right),\label{T1}
\end{align}
where $f_{\tau}=f\tau$.

The average power of the multiuser interference due to the imperfect CSIT is
\begin{align}  \mathbb{E}\left[\left\lVert\tilde{\mathbf{G}}^{\text{T}}\mathbf{P}\mathbf{s}\right\rVert^2\right]&=
\frac{1}{\tau^2}\text{tr}\left(\mathbb{E}\left[\tilde{\mathbf{G}}^*\tilde{\mathbf{G}}^{\text{T}}\right]\mathbf{P}\mathbf{P}^{H}\right).\label{T2}
\end{align}
Constructing the Lagrangian, we obtain
\begin{align}
    \mathcal{L}(\mathbf{P},&f,\lambda)=\nonumber\\
    &K-f_{\tau}^{-1}\left(\text{tr}\left(\hat{\mathbf{G}}^{\T}\mathbf{P}\right)+\text{tr}\left(\mathbf{P}^{H}\hat{\mathbf{G}}^{*}\right)\right)+f^{-2}\text{tr}\left(\mathbf{R}_\mathbf{n}\right)\nonumber\\
    &+f_{\tau}^{-2}\left(\text{tr}\left(\mathbf{P}^{H}\hat{\mathbf{G}}^{*}\hat{\mathbf{G}}^{\text{T}}\mathbf{P}\right)+\text{tr}\left(\mathbf{P}^{H}\E\left[\tilde{\mathbf{G}}^{*}\tilde{\mathbf{G}}^{\T}\right]\mathbf{P}\right)\right)\nonumber\\ 
    &+\frac{1}{\tau^2}\text{tr}\left(\mathbb{E}\left[\tilde{\mathbf{G}}^*\tilde{\mathbf{G}^{\text{T}}}\right]\mathbf{P}\mathbf{P}^{H}\right)+\lambda\left(\text{tr}\left(\mathbf{P}\mathbf{P}^H\right)-P_t\right).\label{Lagrangian of the robust MMSE}
\end{align}
To solve \eqref{Lagrangian of the robust MMSE}, we first compute the derivatives, resulting in
\begin{align}
\frac{\partial\mathcal{L}\left(\mathbf{P},f,\lambda\right)}{\partial\mathbf{P}^{*}}=&-f^{-1}_{\tau}\hat{\mathbf{G}}^{*}+f^{-2}_{\tau}\left(\hat{\mathbf{G}}^{*}\hat{\mathbf{G}}^{\text{T}}+\mathbb{E}\left[\tilde{\mathbf{G}}^{*}\tilde{\mathbf{G}}^{\text{T}}\right]\right)\mathbf{P}\nonumber\\
    &+\frac{1}{\tau^{2}}\mathbb{E}\left[\tilde{\mathbf{G}}^{*}\tilde{\mathbf{G}}^{\text{T}}\right]\mathbf{P}+\lambda\mathbf{P}.\label{Lagrangian precoder}
\end{align}
\begin{align}
\frac{\partial\mathcal{L}\left(\mathbf{P},f,\lambda\right)}{\partial f}=&f^{-2}_{\tau}\left(\text{tr}\left(\hat{\mathbf{G}}^{\T}\mathbf{P}\right)+\text{tr}\left(\mathbf{P}^{H}\tilde{\mathbf{G}}^{*}\right)\right)-\frac{2}{f^3}\text{tr}\left(\mathbf{R}_{\mathbf{n}}\right)\nonumber\\
    &-\frac{2}{\tau^2f^3}\text{tr}\left(\mathbf{P}^H\E\left[\tilde{\mathbf{G}}^{*}\tilde{\mathbf{G}}^{\T}\right]\mathbf{P}\right)\nonumber\\
    &-\frac{2}{\tau^2f^3}\text{tr}\left(\mathbf{P}^{H}\hat{\mathbf{G}}^{*}\hat{\mathbf{G}}^{\text{T}}\mathbf{P}\right).\label{lambda}
\end{align}
Equating \eqref{Lagrangian precoder} to zero and then solving for $f_{\tau}\hat{\mathbf{G}}^{*}$,  
we get
\begin{align}
f_{\tau}\hat{\mathbf{G}}^{*}=&\underbrace{\left(\hat{\mathbf{G}}^{*}\hat{\mathbf{G}}^{\text{T}}+\mathbb{E}\left[\tilde{\mathbf{G}}^{*}\tilde{\mathbf{G}}^{\text{T}}\right]+f^2\mathbb{E}\left[\tilde{\mathbf{G}}^{*}\tilde{\mathbf{G}}^{\text{T}}\right]+\lambda f^2_{\tau}\mathbf{I}\right)}_{\mathbf{P}^{\left(i\right)}}\mathbf{P}.\label{equation for P}
\end{align}
Then, we can isolate $\mathbf{P}$ to obtain the robust precoder. This concludes the proof.
\end{IEEEproof}

\begin{proposition}\label{prop:robust2}
    The parameter $\lambda$ is computed as
    \begin{equation}
        \lambda=\frac{\textup{tr}\left(\mathbf{R}_{\mathbf{n}}\right)}{f^2P_t}-\frac{\textup{tr}\left(\mathbf{P}^{H}\E\left[\tilde{\mathbf{G}}^{*}\tilde{\mathbf{G}}^{\text{T}}\right]\mathbf{P}\right)}{\tau^2P_t}.\label{solution for lambda}
    \end{equation}
\end{proposition}
\begin{IEEEproof}
    Equating \eqref{lambda} to zero and rearranging terms, we get
    \begin{align}
        f_{\tau}\text{tr}\left(\hat{\mathbf{G}}^{*}\mathbf{P}^{H}\right)=&\text{tr}\left(\mathbf{P}^{H}\hat{\mathbf{G}}^{*}\hat{\mathbf{G}}^{\T}\mathbf{P}\right)+\text{tr}\left(\mathbf{P}^{H}\E\left[\tilde{\mathbf{G}}^{*}\tilde{\mathbf{G}}^{\T}\right]\mathbf{P}\right)\nonumber\\
        &\tau^2\text{tr}\left(\mathbf{R}_{\mathbf{n}}\right).\label{first eq for lambda}
    \end{align}
    Multiplying the right-hand side of \eqref{equation for P} by $\mathbf{P}^{H}$ and taking the trace results in
    \begin{align}  f_{\tau}\text{tr}&\left(\hat{\mathbf{G}}^{*}\mathbf{P}^{H}\right)=\nonumber\\
    &\text{tr}\left(\left(\hat{\mathbf{G}}^{*}\hat{\mathbf{G}}^{\text{T}}+\left(1+f^2\right)\mathbb{E}\left[\tilde{\mathbf{G}}^{*}\tilde{\mathbf{G}}^{\text{T}}\right]+\lambda f^2_{\tau}\mathbf{I}\right)\mathbf{P}\mathbf{P}^{H}\right).\label{second eq for lambda}
    \end{align}
    Equating  \eqref{first eq for lambda} and \eqref{second eq for lambda} gives
    \begin{equation}
        \tau^2\text{tr}\left(\mathbf{R}_\mathbf{n}\right)=f^2\text{tr}\left(\E\left[\tilde{\mathbf{G}}^{*}\Tilde{\mathbf{G}}^{\T}\right]\right)+\lambda\tau^2f^2\underbrace{\text{tr}\left(\mathbf{P}\mathbf{P}^{H}\right)}_{P_t}\label{last equation for lambda}
    \end{equation}
    Solving \eqref{last equation for lambda} for $\lambda$ concludes the proof.
\end{IEEEproof}
\begin{proposition}\label{prop:robust3}
    The parameter $f$ is given by
    \begin{equation}
    f=\frac{1}{\tau}\sqrt{\frac{P_t}{\text{tr}\left(\bar{\mathbf{P}}\bar{\mathbf{P}}^{H}\right)}}\label{power scaling factor}
\end{equation} 
\end{proposition}
\begin{IEEEproof}
  Substituting the robust precoder into the power constraint results in  
  \begin{align}
P_{t}=&\text{tr}\left(\mathbf{P}^{\left(\text{r}\right)}\mathbf{P}^{\left(\text{r}\right)^{H}}\right)\nonumber\\
    =&f^2\tau^2\textup{tr}\left(\bar{\mathbf{P}}\bar{\mathbf{P}}^{H}\right).
\end{align}
Solving for $f$ concludes the proof.
\end{IEEEproof}
Note that $\mathbf{P}$ depends on $\lambda$ and vice-versa. To obtain both quantities, we employ an alternating optimization framework, where one of the variables is fixed to compute the other. We begin the computations with the conventional MMSE as the initial state. With this initial precoder, we update the parameter $\lambda$. Algorithm \ref{alg:Robust MMSE} summarizes these steps.

{
\begin{algorithm}[H]
	\caption{Robust MMSE}
	\label{alg:Robust MMSE}
    \begin{algorithmic}[1]
    \Statex \textbf{Input:}  $P_t$, $\sigma_n^2$ ,$\sigma_e^2$, $i_t$
    \Statex \textbf{Output:} $\mathbf{P}^{\left(\text{r}\right)}$
\State 
$\boldsymbol{\Theta} \leftarrow \mathbb{E}\left[\tilde{\mathbf{G}}^{*}\tilde{\mathbf{G}}^{\text{T}}\right]=\sigma_e^2\mathbf{I}$
\State $\bar{\mathbf{P}}\left[0\right]\leftarrow\left(\hat{\mathbf{G}}^{*}\hat{\mathbf{G}}^{\text{T}}+\frac{K\sigma_n^2}{P_t}\mathbf{I}\right)^{-1}\hat{\mathbf{G}}^{*}$
\State
$f\left[0\right]\leftarrow\sqrt{\frac{P_t}{\text{tr}\left(\bar{\mathbf{P}}\left[0\right]\bar{\mathbf{P}}^H\left[0\right]\right)}}$
\State
$\mathbf{P}\left[0\right]\leftarrow f\left[0\right]\bar{\mathbf{P}}\left[0\right]$
\State
$\lambda\left[0\right]\leftarrow\frac{\text{tr}\left(\mathbf{R}_\mathbf{n}\right)}{f^2\left[0\right] P_t}-\frac{\text{tr}\left(\mathbf{P}\left[0\right]\mathbf{P}^{H}\left[0\right]\mathbb{E}\left[\tilde{\mathbf{G}}^{*}\tilde{\mathbf{G}}^{\text{T}}\right]\right)}{P_t}$
\For{$i=1:i_t$}
\State
$\bar{\mathbf{P}}\left[i\right] \leftarrow(\hat{\mathbf{G}}^{*}\hat{\mathbf{G}}^{\text{T}}+\left(1+f^2\left[i-1\right]\right)\mathbb{E}\left[\tilde{\mathbf{G}}^{*}\tilde{\mathbf{G}}^{\text{T}}\right]$
\hspace{1em}$+\lambda\left[i-1\right] f^2\left[i-1\right]\mathbf{I} )^{-1}\hat{\mathbf{G}}^{*}$
\State
$f\left[i\right]\leftarrow\sqrt{\frac{P_t}{\text{tr}\left(\bar{\mathbf{P}}\left[i\right]\bar{\mathbf{P}}^H\left[i\right]\right)}}$
\State 
$\mathbf{P}\left[i\right]\leftarrow f\left[i\right]\bar{\mathbf{P}}\left[i\right]$
\State $\lambda\left[i\right]\leftarrow\frac{\text{tr}\left(\mathbf{R}_\mathbf{n}\right)}{f^2\left[i\right] P_t}-\frac{\text{tr}\left(\mathbf{P}\left[i\right]\mathbf{P}^{H}\left[i\right]\mathbb{E}\left[\tilde{\mathbf{G}}^{*}\tilde{\mathbf{G}}^{\text{T}}\right]\right)}{P_t}$
\EndFor
\State \textbf{end for}
\State
$\mathbf{P}^{\left(\text{r}\right)}\leftarrow f\left[i_t\right]\bar{\mathbf{P}}\left[i_t\right]$
\Statex \Return $\mathbf{P}^{\left(\text{r}\right)}$
\end{algorithmic}\label{alg1 robust MMSE}
\end{algorithm}
}

\section{Robust Precoders}

To reduce the signaling load of the system and the computational complexity of NW precoders, we propose user-centric cluster-based precoders. To this end, clusters of APs and users are formed. These clusters are defined based on the large-scale channel coefficients given by $\zeta_{k,n}$. The motivation behind this selection scheme is that only small subsets of APs transmit the most relevant signals for reception. The benefit is that we discard the APs whose processing is cost-ineffective and reduce the signaling load.

\subsection{AP selection}
To reduce the signaling load, AP selection is performed, so that each user is served only by a subset of APs. The AP selection process is carried out by taking the largest large-scale fading coefficients in the channel. For instance, let us denote by $L$ the number of selected APs. Then, for an arbitrary user, say $k$-th user, the $L$ APs with the largest large-scale fading coefficient are selected and gathered in the set $\mathcal{U}_k$. In this sense, we employ the equivalent channel estimate $\bar{\textbf{G}}^{\text{T}}=\left[\mathbf{\bar{g}}_1,\mathbf{\bar{g}}_2,\cdots,\mathbf{\bar{g}}_k\right]^{\text{T}}\in \mathbb{C}^{K \times N}$, which is a sparse matrix with the $(n,k)$-th element as
\begin{equation}
    \bar{\text{g}}_{k,n}=\begin{cases}
\hat{g}_{k,n},&n\in \mathcal{U}_k,\\
0, &\text{otherwise.}
\end{cases}
\label{sparse effective channel}
\end{equation}

\subsection{Sparse robust MMSE}
We employ the equivalent channel estimate $\bar{\mathbf{G}}^{\T}$ to find a robust precoder that reduces the signaling load. It follows that the sparse robust MMSE (MMSE-RB-SP) is 
\begin{equation}    \mathbf{P}^{\left(\text{sp}\right)}=f_{\tau}{\bar{\mathbf{P}}^{\left(\text{sp}\right)}},
\end{equation}
where
\begin{equation}
    \bar{\mathbf{P}}^{\left(\text{sp}\right)}=\left(\bar{\mathbf{G}}^{*}\bar{\mathbf{G}}^{\text{T}}+\left(1+f^2\right)\mathbb{E}\left[\tilde{\mathbf{G}}^{*}\tilde{\mathbf{G}}^{\text{T}}\right]+\lambda f^2_{\tau}\mathbf{I}\right)^{-1}\bar{\mathbf{G}}^{*}.\label{p_bar robust sp mmse}
\end{equation}
Then, we resort to the alternating optimization framework described in Algorithm \ref{alg1 robust MMSE} by substituting \eqref{p_bar robust sp mmse} in \eqref{solution for lambda} to find the parameter $\lambda$. Once the parameter $\lambda$ is onbtained, we update the precoder in \eqref{p_bar robust sp mmse}. 

\subsection{Cluster-based robust MMSE}
One major disadvantage of the MMSE-RB-SP is that it requires the inversion of multiple matrices, which results in high computational cost. To tackle this problem, we group the users in $K$ clusters. Each cluster computes one column of the precoding matrix by employing a channel matrix with reduced dimensions thereby bringing down the computational cost.  

Denote by $\mathcal{U}_k$ the cluster of users formed to compute the precoder of the $k$-th user, i.e., $\mathbf{p}_k$. User $i$ is included in  $\mathcal{U}_k$ if it shares at least $N_a$ antennas with user $k$, where $N_a$ is a prefixed parameter. In other words, there are $N_a$ antennas providing service to both users, user $i$ and user $k$. With $\mathcal{U}_k$ defined, we have the reduced dimension matrix of $\hat{\textbf{G}}$ as $\hat{\textbf{G}}_k \in\mathbb{C}^{\lvert\mathcal{U}_k\rvert\times K}$. For this purpose, we define $\mathbf{U}_{k}\in\mathbb{R}^{\lvert\mathcal{U}_k\rvert \times K}$, which can be interpreted as a user selection matrix. The first row of $\mathbf{U}_{k}$  is 
\begin{equation}
\mathbf{u}_{1,k}=[\underbrace{0,\cdots,}_{i-1~~\textrm{terms}}1,\cdots,0] \in\mathbb{R}^{\lvert\mathcal{U}_k\rvert}, 
\end{equation}
where $i=\min\limits_{r\in\mathcal{U}_k}r$. Similarly, the second row $\mathbf{u}_{2,k}$ a one at the $l$-th position, where $l$ is the second lowest index in $\mathcal{U}_k$. All the other coefficients of the second row are equal to zero. We similarly obtain the following rows of $\mathbf{U}_k$. Thus, the reduced matrix is $\hat{\textbf{G}}^{\text{T}}_k=\mathbf{U}_k\hat{\textbf{G}}^{\text{T}}$. 

We denote the robust precoder that employs a channel matrix with reduced dimensions to compute its columns as  MMSE-RB-RD. To find the MMSE-RB-RD, we first solve the following optimization problem
\begin{align}
    \left\{\mathbf{P}',f_k\right\}= &\text{argmin}~~\underbrace{\mathbb{E}\left[\lVert\mathbf{s}_k-f_k^{-1}\mathbf{y}_k\rVert^2\right]}_{T_1}+\mathbb{E}\left[\lVert\boldsymbol{\Delta}_k\rVert^2\right]\nonumber\\    
    &\text{subject to}~~\mathbb{E}\left[\lVert\mathbf{x}_k\rVert^2\right]=\text{tr}\left(\mathbf{P}'\mathbf{P}'^H\right)=\frac{\lvert\mathcal{U}_k\rvert
    P_t}{K},\label{minimization problem clustered RB MMSE}
\end{align}
Following Proposition~\ref{prop:robust1}, the solution to \eqref{minimization problem clustered RB MMSE} is 
\begin{equation}    
\mathbf{P}'=f_{\tau}{\bar{\mathbf{P}}}',
\end{equation}
where
\begin{equation}    {\bar{\mathbf{P}}}'=\left(\hat{\mathbf{G}}^{*}_k\hat{\mathbf{G}}^{\text{T}}_k+\left(1+f^2\right)\mathbb{E}\left[\tilde{\mathbf{G}}^{*}_k\tilde{\mathbf{G}}_k^{\text{T}}\right]+\lambda_k f^2_{\tau}\mathbf{I}\right)^{-1}\hat{\mathbf{G}}_k^{*},\label{p_bar robust mmse2}
\end{equation}
Furthermore, using Proposition ~\ref{prop:robust2} and Proposition ~\ref{prop:robust3}, we obtain
\begin{equation}
\lambda_k=\frac{K\textup{tr}\left(\mathbf{R}_{\mathbf{n},k}\right)}{f^2 \lvert\mathcal{U}_k\rvert P_t}-\frac{K\textup{tr}\left(\mathbf{P}'^{H}\E\left[\tilde{\mathbf{G}}^{*}_k\tilde{\mathbf{G}}_k^{\text{T}}\right]\mathbf{P}'\right)}{\tau^2 \lvert\mathcal{U}_k\rvert P_t},\label{solution for lambda clustered}
\end{equation}
\begin{equation}
    f=\frac{1}{\tau}\sqrt{\frac{\lvert\mathcal{U}_k\rvert P_t}{K\text{tr}\left(\bar{\mathbf{P}}'\bar{\mathbf{P}}'^{H}\right)}}.\label{power scaling factor2}
\end{equation} 
Note that $\mathcal{U}_k$ is associated to decoding of symbol $s_k$ but has reduced dimensions. Therefore, we need an index mapping to find the correct precoder for $s_k$. Since $s_k$ will be decoded, we find the index $q$, so that $\mathbf{u}_{q,k}$ contains a one in its $k$-th entry. It follows that the $q$-th column of $\mathbf{P}'$ should be employed employed in $\mathbf{P}^{\left(\text{MMSE-RB-RD}\right)}= [\mathbf{p}_1^{'}\ldots \mathbf{p}_k^{'} \ldots \mathbf{p}_{K}^{'} ]$ i.e.,
\begin{equation}
    \mathbf{p}_k^{'}=\left[\mathbf{P}_{k}^{'}\right]_q.
\end{equation}

\section{Ergodic Sum-Rate}
To evaluate the performance of the proposed robust precoders, we employ the ergodic sum-rate (ESR) defined as
\begin{equation}
    S_r=\E\left[\sum_{k=1}^{K}\bar{R}_k\right],
\end{equation}
where $\bar{R}_k=\mathbb{E}\left[R_k|\hat{\mathbf{G}}\right]$ is the average rate and $R_k$ is the instantaneous rate of the $k$-th user. The rate $\bar{R}_k$ averages out the effects of the imperfect CSIT. Considering Gaussian codebooks, the instantaneous rate is 
\begin{equation}
    R_k=\log_2\left(1+\gamma_k\right),\label{general instantaneous rate}
\end{equation}
where $\gamma_k$ is the signal-to-interference-plus-noise ratio (SINR) at user $k$. 

To obtain $\gamma_k$, we compute the mean power at the $k$-th receiver as
\begin{align}
    \E\left[\lvert y_k\rvert^2\right]=&\mathbb{E}\left[y_k^*y_k\right]\nonumber\\
    =&\frac{1}{\tau^2}\mathbb{E}\left[ s_k^*s_k\lvert\mathbf{g}^{\text{T}}_k\mathbf{p}_k\rvert^2\right]+\mathbb{E}\left[n_k^* n_k\right]\nonumber\\
    &+\frac{1}{\tau^2}\mathbb{E}\left[ \sum\limits_{\substack{i=1\\i\neq k}}^K s_i^*s_i\lvert\mathbf{g}^{\text{T}}_k\mathbf{p}_i\rvert^2\right].
\end{align} 
Simplifying $\E\left[\lvert y_k\rvert^2\right]$ gives
\begin{align}
    \mathbb{E}\left[\lvert y_k\rvert^2\right]=&\frac{1}{\tau^2}\left\lvert\left(\hat{\mathbf{g}}_k^{\textrm{T}}-\tilde{\mathbf{g}}_k^{\textrm{T}}\right)\mathbf{p}_k\right\rvert^2+\sigma_n^2\nonumber\\
    &+\frac{1}{\tau^2}\sum\limits_{\substack{i=1\\i\neq k}}^K a_i^2\left\lvert\left(\hat{\mathbf{g}}_k^{\textrm{T}}-\tilde{\mathbf{g}}_k^{\textrm{T}}\right)\mathbf{p}_i\right\rvert^2.
\end{align}
Therefore, we have
\begin{equation}
\gamma_k=\frac{\lvert\hat{\mathbf{g}}^{\textrm{T}}_k\mathbf{p}_k\rvert^2}{d_g+\sum\limits_{\substack{i=1\\ i\neq k}}^K \left\lvert\left(\hat{\mathbf{g}}_k^{\textrm{T}}-\tilde{\mathbf{g}}_k^{\textrm{T}}\right)\mathbf{p}_i\right\rvert^2+\tau^2\sigma_n^2}\label{SINR general precoder},
\end{equation}
where $d_g=\lvert\tilde{\mathbf{g}}_k^{\textrm{T}}\mathbf{p}_k\rvert^2-2\Re{\left\{\left(\hat{\mathbf{g}}_k^{\textrm{T}}\mathbf{p}_k\right)^*\left(\tilde{\mathbf{g}}^{\textrm{T}}_k\mathbf{p}_k\right)\right\}}$.
\section{Numerical Experiments}
We evaluate the performance of the proposed robust precoders via numerical experiments. Throughout all experiments, the large scale fading coefficients are set to
\begin{equation}
     \zeta_{k,n}=P_{k,n}\cdot 10^{\frac{\sigma^{\left(\textrm{s}\right)}z_{k,n}}{10}},
 \end{equation}
 where $P_{k,n}$ is the path loss and shadowing effect is included in the scalar $10^{\frac{\sigma^{\left(\textrm{s}\right)}z_{k,n}}{10}}$ with $\sigma^{\left(\textrm{s}\right)}=8$. The random variable $z_{k,n}$ follows Gaussian distribution with zero mean and unit variance. The path loss was obtained by employing the three-slope model: \par\noindent\small
 \begin{align}
     P_{n,k}=\begin{cases}
  -L-35\log_{10}\left(d_{n,k}\right), & \text{$d_{n,k}>d_1$} \\
  -L-15\log_{10}\left(d_1\right)-20\log_{10}\left(d_{n,k}\right), & \text{$d_0< d_{n,k}\leq d_1$}\\
    -L-15\log_{10}\left(d_1\right)-20\log_{10}\left(d_0\right), & \text{otherwise,}
\end{cases}
 \end{align}\normalsize
 where $d_{n,k}$ is the distance between the $n$-th AP and $k$-th user, $d_1=50$ m, $d_0= 10$ m. The attenuation $L$ is \par\noindent\small
 \begin{align}
     L=&46.3+33.9\log_{10}\left(f\right)-13.82\log_{10}\left(h_{\textrm{AP}}\right)\nonumber\\
     &-\left(1.1\log_{10}\left(f\right)-0.7\right)h_u+\left(1.56\log_{10}\left(f\right)-0.8\right),
 \end{align}\normalsize
 where $h_{\textrm{AP}}=15$ m and $h_{u}=1.65$ m are the positions of the APs and UEs above the ground, respectively. The carrier frequency was set to $f= 1900$ MHz. The noise variance is 
 \begin{equation}
     \sigma_n^2=T_o k_B B N_f,
 \end{equation}
 where $T_o=290$ K is the noise temperature, $k_B=1.381\times 10^{-23}$ J/K is the Boltzmann constant, $B=50$ MHz is the bandwidth and $N_f=10$ dB is the noise figure. The signal-to-noise (SNR) is given by
 \begin{equation}
     \text{SNR}=\frac{P_{t}\textrm{Tr}\left(\mathbf{G}^{\text{T}}\mathbf{G}^{*}\right)}{N K \sigma_n^2}.
 \end{equation}\normalsize
In all simulations, we consider a total of 128 APs providing service to 16 users. The APs and the users were randomly distributed. To compute the average rate, we employ $100$ different error matrices $\tilde{\mathbf{G}}$. Moreover, we considered a total of $100$ channel estimates, which results in  $10,000$ trials to compute the ESR.

\begin{figure}[t]
\begin{center}
\includegraphics[width=0.85\columnwidth]{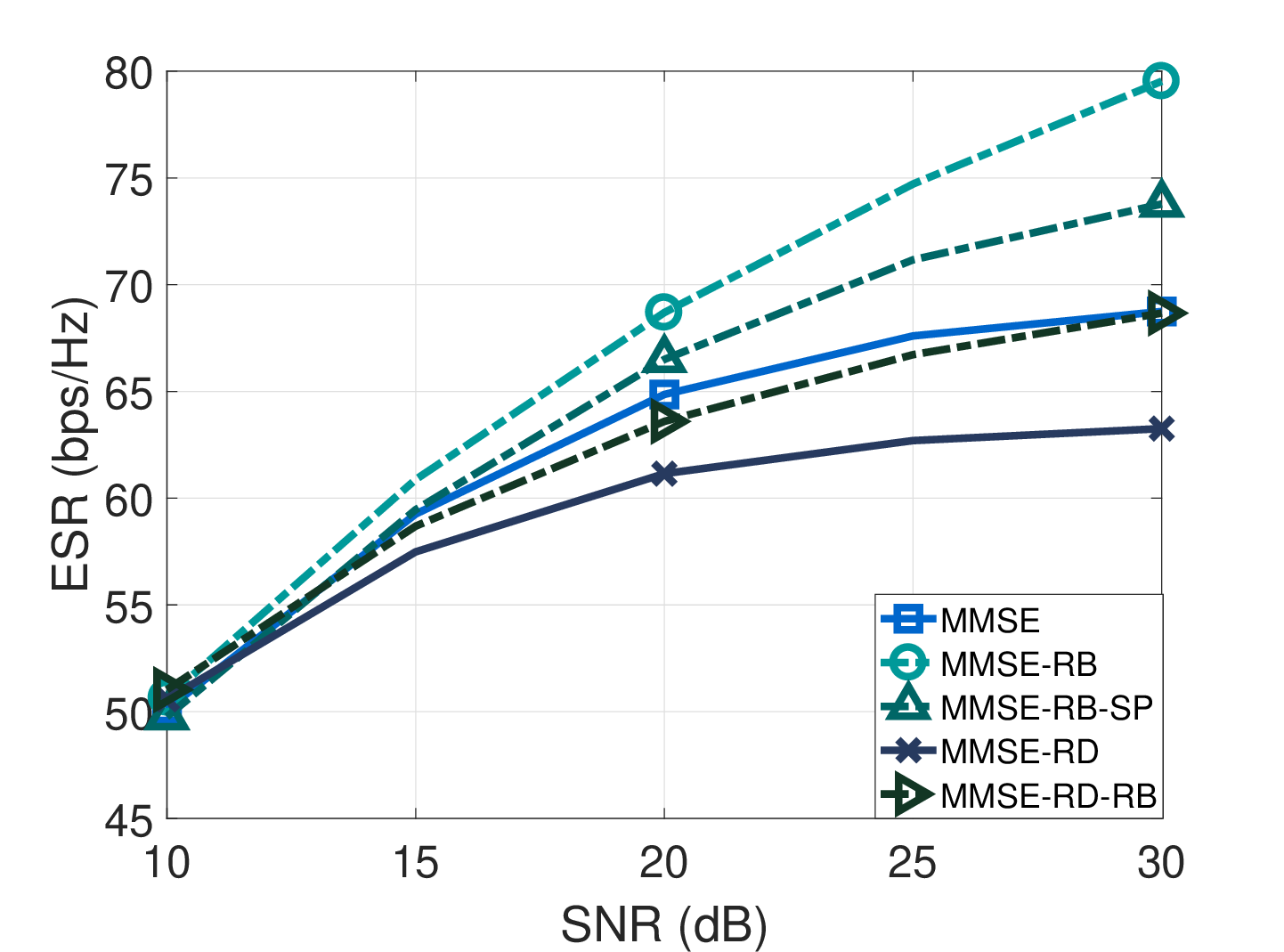}
\vspace{-0.25cm}
\caption{Sum-rate performance of the proposed robust precoders. Here, $M=128$, $K=16$.} 
\label{Fig1}
\vspace{-1.em}
\end{center}
\end{figure}

 Fig. \ref{Fig1} shows the sum-rate performance of the proposed robust precoders compared with the conventional MMSE precoder. The best performance is obtained by robust MMSE (MMSE-RB). The sparse robust MMSE (MMSE-RB-SP) and the robust MMSE with reduced dimensions (MMSE-RB-RD) exhibit a performance loss with respect to the MMSE-RB with the benefit of a lower computational complexity and signaling load.
 
\begin{figure}[t]
\begin{center}
\includegraphics[width=0.85\columnwidth]{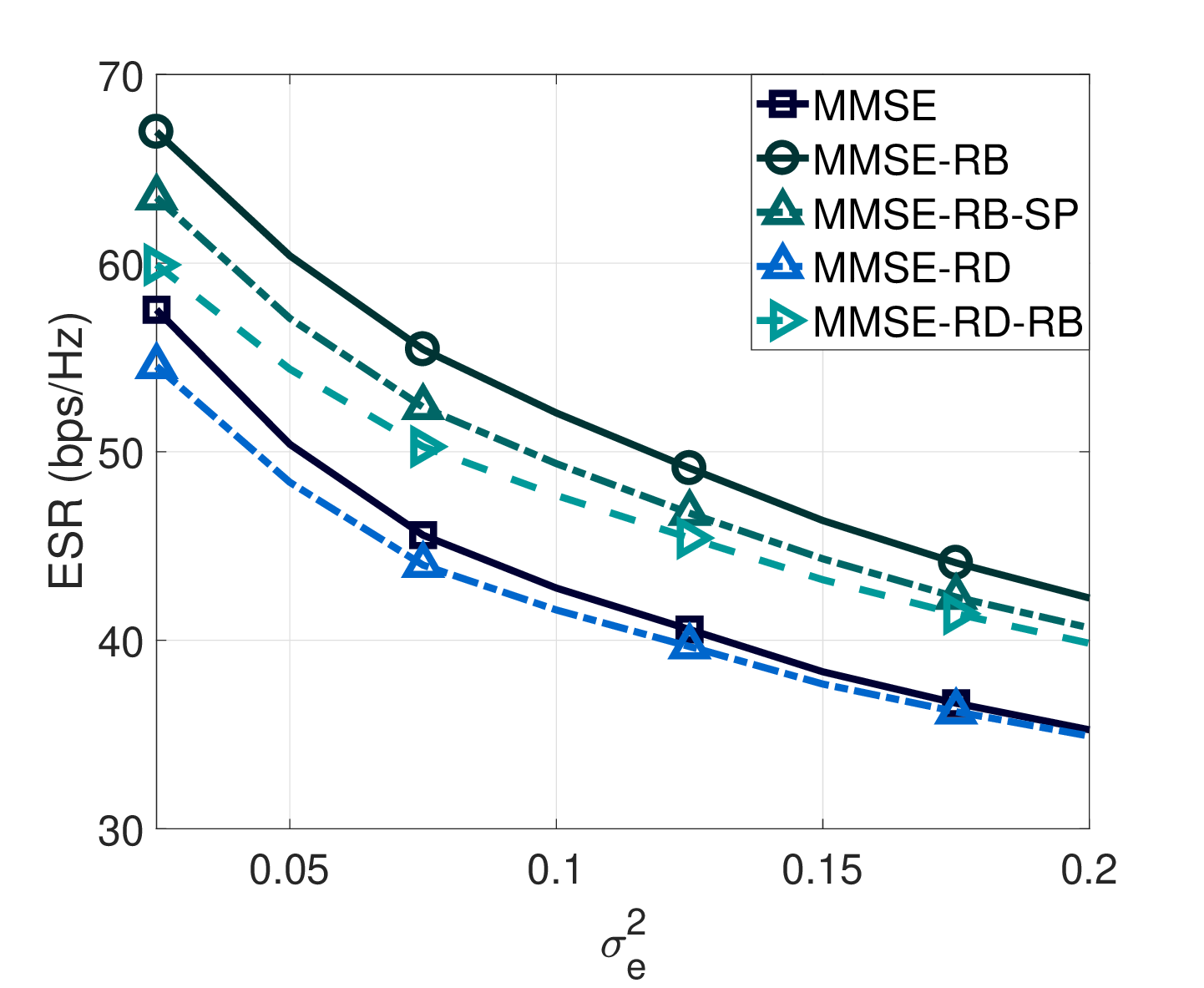}
\vspace{-0.25cm}
\caption{Sum-rate performance vs channel quality of the proposed robust precoders. Here, $M=128$, $K=16$.} 
\label{Fig2}
\vspace{-1.em}
\end{center}
\end{figure}

Next, we assess the sum-rate performance at a SNR of 15 dB for different CSIT qualities (Fig.~\ref{Fig2}). The proposed MMSE-RD-RB outperforms the conventional MMSE-RD. In general, the proposed robust MMSE techniques outperform the conventional MMSE-based precoders. It follows from Fig. \ref{Fig2} that our proposed techniques are effective against CSIT uncertainties.  

\section{Summary}
We developed robust MMSE-based precoding techniques for CF-MIMO system. The proposed techniques show a higher tolerance to the imperfections that arise naturally in the CSIT estimate. The proposed MMSE-RB-SP and MMSE-RB-RD have lower computational complexity and signaling load. This allows the implementation of scalable systems but at the expense of sum-rate performance.

\bibliographystyle{IEEEtran}
\bibliography{SubsetBib}

\end{document}